\documentclass[aps,prd,groupedaddress,twocolumn]{revtex4}
\usepackage[parfill]{parskip}    
\usepackage{graphicx}
\usepackage{amssymb}
\usepackage{epstopdf}
\usepackage{color}
\usepackage{float}
\usepackage{ulem}
\usepackage{graphicx}
\usepackage{pdfpages}
\usepackage{mathtools}
\usepackage[colorlinks=true,linkcolor=blue,citecolor=blue,urlcolor=blue,breaklinks]{hyperref}

\usepackage{xcolor} 

\begin{document}

\title{Spontaneous scalarization in Einstein--power--Maxwell--scalar models}

\author{M. Carrasco-H.}
\email{cmikaelasalome@gmail.com}
\affiliation{Departamento de F\'isica, Colegio de Ciencias e Ingenier\'ia, Universidad San Francisco de Quito,  Quito 170901, Ecuador\\}

\author{N. M.  Santos}
\email{nunomoreirasantos@tecnico.ulisboa.pt}
\affiliation{Departamento de Física, Instituto Superior Técnico - IST, Universidade de Lisboa - UL, Av. Rovisco Pais 1, 1049-001 Lisboa, Portugal\\
Center for Research and Development in Mathematics and Applications (CIDMA), Department of Mathematics, University of Aveiro, 3810-193 Aveiro, Portugal}

\author{E. Contreras }
\email{econtreras@usfq.edu.ec}
\affiliation{Departamento de F\'isica, Colegio de Ciencias e Ingenier\'ia, Universidad San Francisco de Quito (USFQ),  Quito 170901, Ecuador\\}


\begin{abstract}
We study the spontaneous scalarization of charged black holes in Einstein's gravity minimally coupled to power--Maxwell electrodynamics which, in turn, is non--minimally coupled to a real scalar field. We point out the existence of a specific power for which the scalarized solution is well-behaved, and entropically preferred in comparison to the scalar-free charged black hole solution.
\end{abstract}

\maketitle

\section{Introduction}\label{intro}

Black holes (BHs) play a central role in contemporary theoretical physics and astrophysics. Recent advancements in our understanding of BHs have been greatly influenced by experimental endeavors, such as the detection of gravitational waves by the Laser Interferometer Gravitational-Wave Observatory (LIGO) Scientific Collaboration and Virgo Collaboration \cite{LIGOScientific:2016aoc} and the remarkable imaging of BH shadows by the Event Horizon Telescope (EHT) \cite{EventHorizonTelescope:2019dse}. 

Gravitational-wave astronomy and BH imaging will pave the way for exploring yet-to-be-observed phenomena and offer promising avenues for conducting crucial tests that can either confirm or challenge the predictions of general relativity (GR) and modified gravity theories. One such phenomenon is the well-known spontaneous scalarization of BHs which could lead to the formation of BHs that include ``hair''~\cite{Herdeiro:2015waa} in the form of a non-trivial scalar field. This modification of GR BH solutions challenges the no-hair conjecture \cite{Ruffini:1971bza}, which posits that the only observable properties of a BH should be mass, charge, and angular momentum, regardless of the nature of the matter falling into it.

In recent years, substantial efforts have been made to understand the spontaneous scalarization of BHs. In the pioneering works~\cite{Doneva:2017bvd,Silva:2017uqg} (see also~\cite{Antoniou:2017acq}), it was demonstrated that, within a class of extended scalar--tensor--Gauss-Bonnet theories, new BH solutions emerge from the spontaneous scalarization of Schwarzschild BHs under extreme curvature conditions, in contrast to the conventional spontaneous scalarization of neutron stars, which is triggered by the presence of matter \cite{Damour:1993hw}. This can be seen at the linear level, in which 
Schwarzschild BHs become unstable against scalar perturbations when their size is sufficiently small. Additionally, within this class of models, BHs with scalar hair exist and, depending on the model, can be entropically favored over GR solutions. Some of these scalarized BHs also demonstrate stability against spherical perturbations.  Spontaneous scalarization in scalar--Gauss--Bonnet models has been explored in a variety of contexts in recent years -- see $e.g$ \cite{Minamitsuji:2018xde,Blazquez-Salcedo:2018jnn,Silva:2018qhn,East:2021bqk,Cunha:2019dwb,Herdeiro:2020wei,Macedo:2019sem,Dima:2020yac,Rahimi:2023mgu,Kuroda:2023zbz,Bahamonde:2022chq,Staykov:2022uwq,Doneva:2022yqu,Wang:2020ohb} and~\cite{Doneva:2022ewd} for a review.

In a parallel development of theoretical interest, in particular due to the simplicity of the model, it was shown that charged BHs in Einstein's theory with a Maxwell field non-minimally coupled to a real scalar field, Reissner--Nordstr\"om (RN) BHs also scalarize for sufficiently high charge to mass ratio~\cite{Herdeiro:2018wub}. This opened a new playground, leading to many generalizations as well as detailed theoretical and phenomenological studies~\cite{Myung:2018vug,Myung:2018jvi,Herdeiro:2019yjy,Fernandes:2019rez,Brihaye:2019kvj,Herdeiro:2019oqp,Myung:2019oua,Astefanesei:2019pfq,Konoplya:2019goy,Konoplya:2019fpy,Fernandes:2019kmh,Zou:2019bpt,Herdeiro:2019iwl,Hod:2019ulh,Blazquez-Salcedo:2020nhs,Fernandes:2020gay,Herdeiro:2020iyi,Hod:2020ius,Yu:2020rqi,LuisBlazquez-Salcedo:2020rqp,Myung:2020dqt,Herdeiro:2020htm,Blazquez-Salcedo:2020crd,Myung:2020ctt,Hod:2020cal,Guo:2021zed,Yao:2021zid,Zhang:2021nnn,Xiong:2022ozw,Hod:2022txa,Niu:2022zlf,Jiang:2023gas,Jiang:2023yyn,Guo:2023mda,Hod:2023jdc,Kiorpelidi:2023jjw,Belkhadria:2023ooc}. Thus,  in this case, spontaneous
scalarization of electrovacuum BHs occurs in a model that has no higher curvature corrections, which is an important simplification over the extended scalar--tensor--Gauss-Bonnet models. 

A natural step in exploring the scalarization of electrovacuum BHs is to examine it within the context of Einstein--nonlinear electrodynamics theories. To date, the only study addressing this matter is \cite{Wang:2020ohb}, in which the authors consider a model featuring a non-minimal coupling between a scalar field and the Born-Infeld term \cite{Born:1933qff}. Since nonlinear electrodynamics theories have been a fruitful playground for theoretical experiments, and, since they provide various interesting BH solutions in four and higher dimensions (for instance, regular BHs \cite{Ayon-Beato:1998hmi,Ayon-Beato:1999kuh,Bronnikov:2000vy,Balart:2014cga,Dymnikova:2015hka,Rodrigues:2017yry, Bronnikov:2017sgg}), a wider exploration of spontaneous scalarization within Einstein-non linear electrodynamics theories is appealing.

For one specific class of nonlinear electrodynamics models, referred to as ``power-Maxwell models", the matter source is determined by a Lagrangian expressed as an arbitrary power $n$ of the Maxwell invariant, i.e. $(F_{\mu\nu}F^{\mu\nu})^{n}$ \cite{Hassaine:2007py, Hassaine:2008pw}. When the power of the absolute value of the Maxwell invariant is set to be $d/4$ (where $d$ represents the spacetime dimension), the power--Maxwell action exhibits conformal invariance. However, this conformal invariance is not upheld in higher dimensions, despite the RN solution maintaining it in four dimensions. In light of this, in \cite{Hassaine:2008pw} the authors introduce a relaxation of the conformal condition by considering an arbitrary power of the Maxwell scalar. Within their research, they demonstrate that, for values of $n$ greater than $1/2$ or less than 
$0$, the scalar curvature displays a singularity at the origin. On the other hand, for values of $n$ ranging from $0$ to $1/2$, the scalar curvature diverges at infinity. Notably, when $n$ falls within the range of $(1/2, 3/2)$ with a rational number with an odd denominator, the solution exhibits behavior akin to the standard RN solution in the sense that the charge contribution in the metric decreases more rapidly than the mass contribution. In this work, we study the spontaneous scalarization on an Einstein--power--Maxwell (EPM) system for $n=3/5$.

This work is organized as follows. In the next section, we introduce generalities of spontaneous scalarization driven by a non--minimal coupling between a real  scalar field and a generic matter source. In section \ref{EPM}, we introduce the main aspects of EPM systems we will develop and in section \ref{results} we show and discuss the main results of the work. The last section is dedicated to the final comments and conclusions. 

\section{the model}\label{spontaneousGeneral}

Consider the following action (in this work we will use units such that $4\pi G=c=1$)
\begin{equation}\label{eq:action}
    \mathcal{S}=\frac{1}{4}\int \text{d}^4x~\sqrt{-g} \left[R-2\nabla_{\mu}\phi\nabla^{\mu}\phi-f(\phi)\mathcal{I}\right]\ ,
\end{equation}
where $R$ is the Ricci scalar, $\phi$ is the real scalar field, and $\mathcal{I}$ is a generic matter source non--minimally coupled to the scalar field through the coupling function $f(\phi)$ . Besides, consider a generic static and spherically symmetric metric parameterized as 
\begin{equation}\label{eq:metric}
    \text{d}s^2=-N(r)e^{-2\delta(r)}\text{d}t^2+\frac{\text{d}r^2}{N(r)}+r^2\text{d}\Omega^2\ ,
\end{equation}
where $d\Omega^{2}=d\theta^{2}+\sin^{2}\theta$ and
\begin{equation}
    N(r)=1-\frac{2m(r)}{r}\ ,
\end{equation}
with $m$ the Misner--Sharp mass. The event horizon $r=r_H$ is the root of $N$, i.e. $N(r_H)=0$. As demonstrated in \cite{Herdeiro:2018wub}, spontaneous scalarization occurs whenever the scalar free--solution is unstable under perturbations, which could depend on either the nature of the matter sector $\mathcal{I}$ or the coupling function $f(\phi)$. Indeed, by varying the action in Eq. (\ref{eq:action}) with respect to the scalar field, we arrive at
\begin{equation}
    \Box\phi-f_{,\phi}(\phi)\frac{\mathcal{I}}{4}=0\ ,\label{eq:campo}
\end{equation}
where $\Box=\nabla_{\mu}\nabla^{\mu}$ is the d'Alambertian, and $f_{,\phi}$ is the derivative of $f$ with respect to the scalar field. $f(\phi)$ must allow for the existence of scalar-free solutions (i.e. with $\phi=0$), 
which requires $f(0)_{,\phi}=0$. Now, performing a small perturbation of the scalar field, $\phi=\phi_0+\delta\phi$, with $\phi_0=0$ (free solution), we obtain
\begin{equation}\label{eq:perturbation}
    \left(\Box-f_{,\phi\phi}(0)\frac{\mathcal{I}}{4}\right)\delta\phi=0\ ,
\end{equation}
where $f_{,\phi\phi}$ represents the second derivative of $f$ with respect to the scalar field. Identifying the effective mass as,
\begin{equation}
    \mu_{\rm eff}^2\equiv f_{,\phi\phi}(0)\frac{\mathcal{I}}{4}\ ,
\end{equation}
we observe that for $f_{,\phi\phi}(0)>0$ and $\mathcal{I}<0$ or 
$f_{,\phi\phi}(0)<0$ and $\mathcal{I}>0$, $\mu_{\rm eff}^2<0$ and a tachyonic instability arises, driving the system away from the scalar-free solution. The mechanism of how such an instability evolves dynamically is discussed in \cite{Herdeiro:2018wub}, wherein $\mathcal{I}=F_{\mu\nu}F^{\mu\nu}$, with $F=\text{d}A$ being the Maxwell tensor.

\section{Einstein--power--Maxwell--scalar system}\label{EPM}
%
We will explore the spontaneous scalarization triggered by a power--Maxwell scalar
\begin{equation}\label{matterL}
\mathcal{I}=\left(F_{\mu\nu}F^{\mu\nu}\right)^n\ .
\end{equation}
In this case, the variational principle on (\ref{eq:action}) leads to the Einstein field equations,
\begin{widetext}
\begin{eqnarray}
    && \Box\phi-f_{,\phi}(\phi)\frac{\left(F_{\mu\nu}F^{\mu\nu}\right)^{n}}{4}=0\ ,\\
    &&\frac{1}{\sqrt{-g}}\partial_{\mu}\left[\sqrt{-g}f(\phi) \left(F_{\alpha\beta}F^{\alpha\beta}\right)^{n-1}F^{\mu\nu}\right]=0\ ,\\
    &&G_{\mu\nu}+g_{\mu\nu}(\nabla\phi)^2-2\nabla_{\mu}\phi\nabla_{\nu}\phi+\frac{g_{\mu\nu}}{2}f(\phi)\left(F_{\alpha\beta}F^{\alpha\beta}\right)^n-2f(\phi)n\left(F_{\alpha\beta}F^{\alpha\beta}\right)^{n-1}F_{\mu}^{\ \sigma}F_{\nu\sigma}=0\ ,
\end{eqnarray}
\end{widetext}
with $G_{\mu\nu}$ the Einstein tensor.
By replacing the metric (\ref{eq:metric}) in the above expressions, and $F_{\mu\nu}=\partial_{\mu}A_{\nu}-\partial_{\nu}A_{\mu}$ with $A=V(r)\text{d}t$,  we obtain
\begin{eqnarray}
    && m'=\frac{1}{2}r^2 N\phi^{\prime 2}-2^{n-2}(2n-1)r^2 f(\phi)\left(-e^{2\delta}V^{\prime 2}\right)^n\ ,\quad \quad \label{eq:motion1}\\
&&\delta'+r\phi^{\prime 2}=0\ ,\label{eq:motion2}\\
&& \left[e^{\delta}f(\phi)r^2\left(-2e^{2\delta} V^{\prime 2}\right)^{n-1}V'\right]'=0\ ,\label{eq:motion3}\\
&&\left(e^{-\delta}r^2\phi' N\right)'=2^{n-2}r^2f_{,\phi}(\phi)e^{-\delta}\left(-e^{2\delta}V^{\prime 2}\right)^n\label{eq:motion4}\ ,
\end{eqnarray}
which coincides with the set of equations for the scalarized BH in \cite{Herdeiro:2018wub} when $n\to1$, as expected. At this point, some comments are in order. First, note  that, if the mass function, $m$ is to be an increasing function, $m'$ must be positive everywhere. Now, by examining Eq. (\ref{eq:motion1}), this requirement can be ensured if either $2n-1>0$ and $(-1)^n<0$ or $2n-1<0$ and $(-1)^n>0$. Second, the integration of (\ref{eq:motion3}) is straightforward and can be written as
\begin{equation}\label{eq:potentialPrime}
    V'(r)=e^{-\delta}\left[\frac{(-1)^n Q}{2^{n-1} r^2f(\phi)}\right]^{\frac{1}{2n-1}},
\end{equation}
where $Q$ is identified as the electric charge. The potential in \cite{Herdeiro:2018wub} is recovered when $n\to1$.

The system (\ref{eq:motion1})--(\ref{eq:motion4}) corresponds to a (non-linear) boundary eigenvalue problem, which cannot be solved analytically. However, one can adopt different numerical techniques, with the \textit{shooting method} standing out as the most straightforward choice. This method entails an iterative process wherein initial guess values for the unknowns are adjusted to meet the prescribed conditions at the boundaries. In this model, the set of functions $\{m,\delta,\phi\}$ needs to be numerically solved, for any $r\in(r_H,\infty)$. As we require an asymptotically flat solution, the conditions at infinity are well established: $m\to M$, $\delta\to0$, $\phi \to 0$. Nevertheless, the exact values of $\delta$ and $\phi$ at $r_H$ remain undetermined. To address this problem, let us employ an expansion of the functions in the vicinity of $r_H$ 

%
\begin{equation}\label{eq:expansion}
    \begin{split}
        m(r)&= \frac{r_H}{2}+m_1(r-r_H)+...\ ,\\
        \delta(r)&= \delta_0+\delta_1(r-r_H)+...\ ,\\
        \phi(r)&= \phi_0+\phi_1(r-r_H)+... \ .
    \end{split}
\end{equation}
\begin{widetext}
Replacing the expansions into the equations of motion, one finds
\begin{eqnarray}
    &&\delta_1=-\phi_1^2r_H\ ,\label{eq:delta1}\\
    &&m_1=(-1)^{n+1}2^{n-2}(2n-1)f(\phi_0)r_H^2\left[\frac{(-1)^n Q}{2^{n-1}n r^2f(\phi_0)}\right]^{\frac{2n}{2n-1}}\ ,\label{eq:m1}\\
    &&\phi_1=\frac{(-1)^n 2^{n-2}nr_H f_{,\phi}}{n+(2n-1)Q\left[\frac{(-1)^n Q}{2^{n-1}n r^2f(\phi)}\right]^{\frac{2n}{2n-1}}}\left[\frac{(-1)^n Q}{2^{n-1}n r_H^2f(\phi)}\right]^{\frac{2n}{2n-1}}\ ,\label{eq:phi1}
\end{eqnarray}
from where we have identify our unknowns: $\delta(r_H)=\delta_0$ and $\phi(r_H)=\phi_0$. Furthermore, note that our system remains invariant under $\delta\to\delta+\Tilde{\delta}$, where $\Tilde{\delta}$ is a constant, so we can initially set $\delta_0=0$ and use the symmetry to recover the physical solutions afterward. Consequently, we reduce the number of unknown parameters to just one, $\phi_0$, which will serve as the \textit{shooting parameter}. Fixing $\{r_H,Q\}$, we choose a guess value for the shooting parameter, integrate the equations of motion and repeat the previous steps, updating the guess value for $\phi_0$ until $\phi$ vanishes at (numerical) spatial infinity. The latter is the shooting condition.

The solutions must satisfy the following virial identity \cite{Herdeiro:2018wub}
\begin{equation}\label{eq:virialPW}
        \int_{r_H}^{\infty}\text{d}r \left\{e^{-\delta}r^2 \phi^{\prime 2}\left[1+\frac{2r_H}{r}\left(\frac{m}{r}-1\right)\right]\right\}=-\frac{1}{2}\int_{r_H}^{\infty}\text{d}r \left\{r^2 e^{-\delta-\alpha\phi^2}\left(-2e^{2\delta}V^{\prime 2}\right)^n\left(3-2n-\frac{2r_H}{r}\right)\right\}\ ,
\end{equation}
\end{widetext}
from where we observe that, as the left-hand side of (\ref{eq:virialPW}) is consistently positive, a nontrivial scalar field can only be supported for $Q\ne0$.

To determine a specific value of $n$, let us take a closer look at the classical solution for the power Maxwell model, whose mass function reads \cite{Rincon:2021PW,Hassaine:2008pw}
\begin{equation}
    m(r)=M-\frac{\left(2Q^2\right)^{n}}{45r^{\beta-1}}\ ,
\end{equation}
where $M$ and $Q$ are the mass and the charge of the BH and $\beta$ is related to the power $n$ by
\begin{equation}\label{eq:beta}
    \beta=\frac{2}{2n-1}\ .
\end{equation}
It is worth noticing that, although $\beta$ could be arbitrary, the simplest choice is to consider it as a natural number, in which case $n$ must take the values shown in \autoref{tab:nValues}. All values therein satisfy $2n-1>0$. Now, from the whole list, the only acceptable values for $n$ are those leading to $(-1)^{n}\in\mathbb{R}$, namely $n=\{1,2/3,3/5\}$. However, note that $(-1)^n>0$ for $n=2/3$, which violates the requirement for an increasing mass function, as discussed previously. Besides, $n=1$ corresponds to the RN case covered in \cite{Herdeiro:2018wub}, so we will not discuss this case here and the only option is $n=3/5$ \footnote{Note that $(-1)^{3/5}$ is a multivalued complex function. However, in this work we are considering the real branch which leads to $(-1)^{3/5}=-1$}.
\begin{table}[h!]
    \centering
    \begin{tabular}{|c||c|c|c|c|c|c|c|c|c|c|}
        \toprule
         $\beta$ & 1 & 2 & 3 & 4 & 5 & 6 & 7 & 8 & 9 & 10 \\
         \colrule
         $n$ & $\frac{3}{2}$ &1 & $\frac{5}{6}$ & $\frac{3}{4}$ & $\frac{7}{10}$ & $\frac{2}{3}$ & $\frac{9}{14}$ & $\frac{5}{8}$ & $\frac{11}{18}$ & $\frac{3}{5}$ \\
        \botrule
    \end{tabular}
    \caption{Corresponding values of $n$ for the first ten natural values of $\beta$.}
    \label{tab:nValues}
\end{table}

To fix the coupling function, we observe the Bekenstein-type identities, that read $\phi f_{,\phi}>0$ and $f_{,\phi\phi}>0$ \cite{Herdeiro:2018wub}; we then take 
\begin{equation}\label{coupling}
    f(\phi)=e^{-\alpha\phi^2}\ ,
\end{equation}
with $\alpha<0$, as the coupling function.

\section{Domain of existence and numerical solution}\label{results}

\begin{figure}[h!]
    \centering
    \includegraphics[width=\columnwidth]{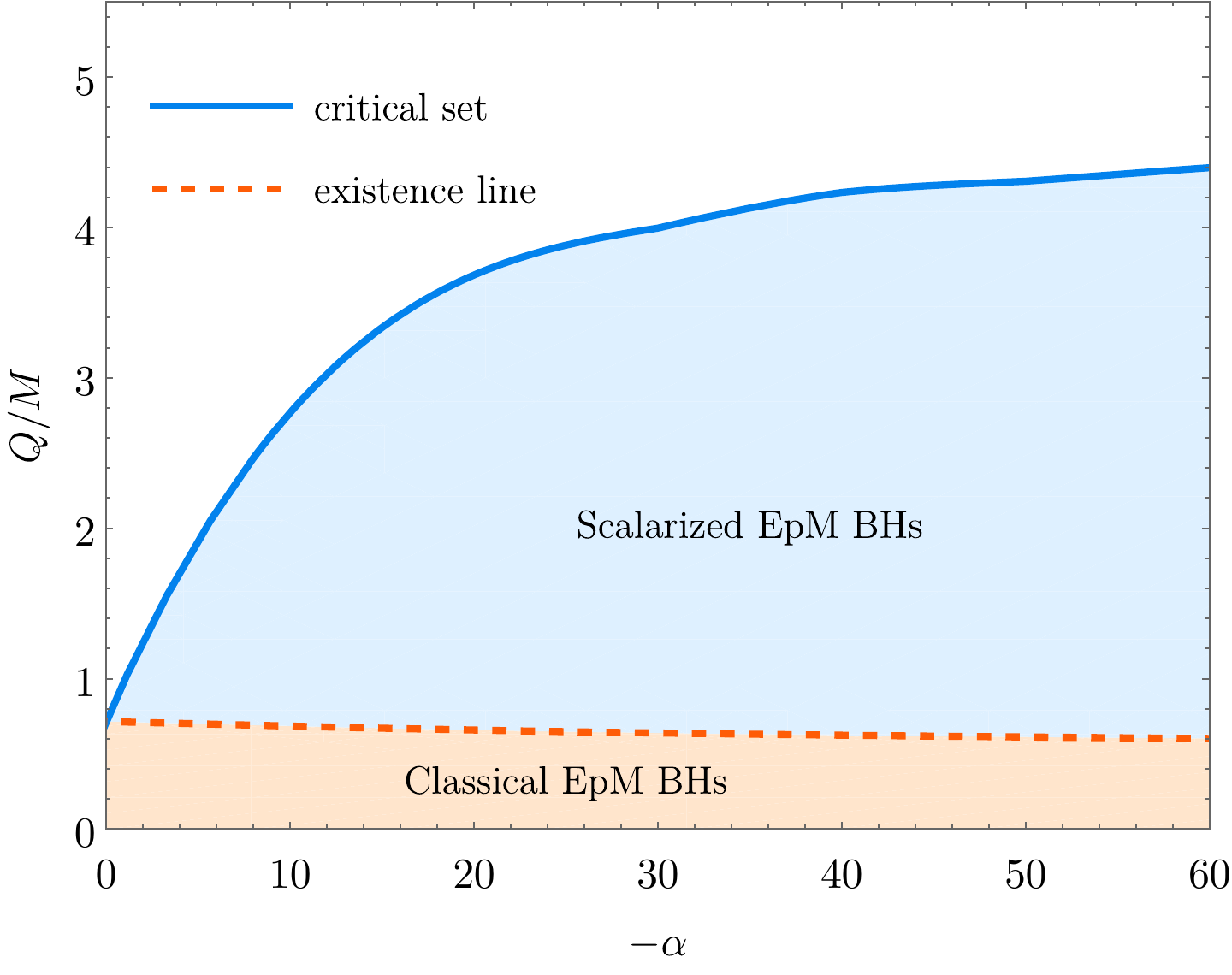}
    \caption{Part of the domain of existence of spherical scalarized EPM BHs with $n=3/5$ in the $(\alpha, q)$--plane. }
    \label{fig:bifurcation}
\end{figure}

The scalar-free EPM solution with $n=3/5$ is described by $\phi=0$ and (\ref{eq:metric}), with $\delta=0$ and \cite{Rincon:2021PW,Hassaine:2008pw} 
\begin{equation}
    N(r)=1-\frac{2M}{r}+\frac{2\left(2Q^2\right)^{3/5}}{45r^{10}}\ .
\end{equation}
One can decompose the scalar field perturbation as
\begin{equation}\label{eq:decomposition}
    \delta \phi(r,\theta,\varphi)=\sum_{\ell,m}Y_{\ell,m}(\theta,\varphi)U_{\ell}(r)\ .
\end{equation}
Then, replacing (\ref{eq:decomposition}) and (\ref{matterL}) in the scalar field equation (\ref{eq:perturbation}), we obtain
\begin{equation}\label{eom-pert}
    \frac{e^{\delta}}{r^2}\frac{d}{dr}\left(\frac{r^2 N}{e^{\delta}}\frac{dU_{\ell}}{dr}\right)-\left[\frac{\ell(\ell+1)}{r^2}+\mu_{\rm eff}^2\right]U_{\ell}=0\ ,
\end{equation}
with $\mu_{\rm  eff}^2=4\alpha \left(\frac{5Q}{3r^2}\right)^6<0$. The solution of (\ref{eom-pert}) defines the lower bound, known as the \textit{existence line} (orange dotted line in \autoref{fig:bifurcation}), for the domain of existence of the scalarized solutions (blue area in \autoref{fig:bifurcation}). For the spherical configuration $\ell=0$ and $m=0$, $U_0$ depends solely on the generic parameters $q=Q/M$ and $\alpha$, so obtaining the points of the existence line is reduced to studying the zeros of $U_0$ as $r\to\infty$. The domain of existence can be obtained through numerical iteration by fixing $\alpha$ and $Q$ and varying $r_H$. For each $r_H$, the equations of motion \eqref{eq:motion1}--\eqref{eq:motion4} are solved, and the initial guess for $\phi_0$ is the value obtained for a neighbor solution. Each $\alpha$--branch ends at \textit{critical set} (blue line in \autoref{fig:bifurcation}), defined by a vanishing horizon area. The scalarized solutions exhibit a virial value of the order of $10^{-6}$.  

\begin{figure}[h!]
    \centering
    \includegraphics[width=8.5cm]{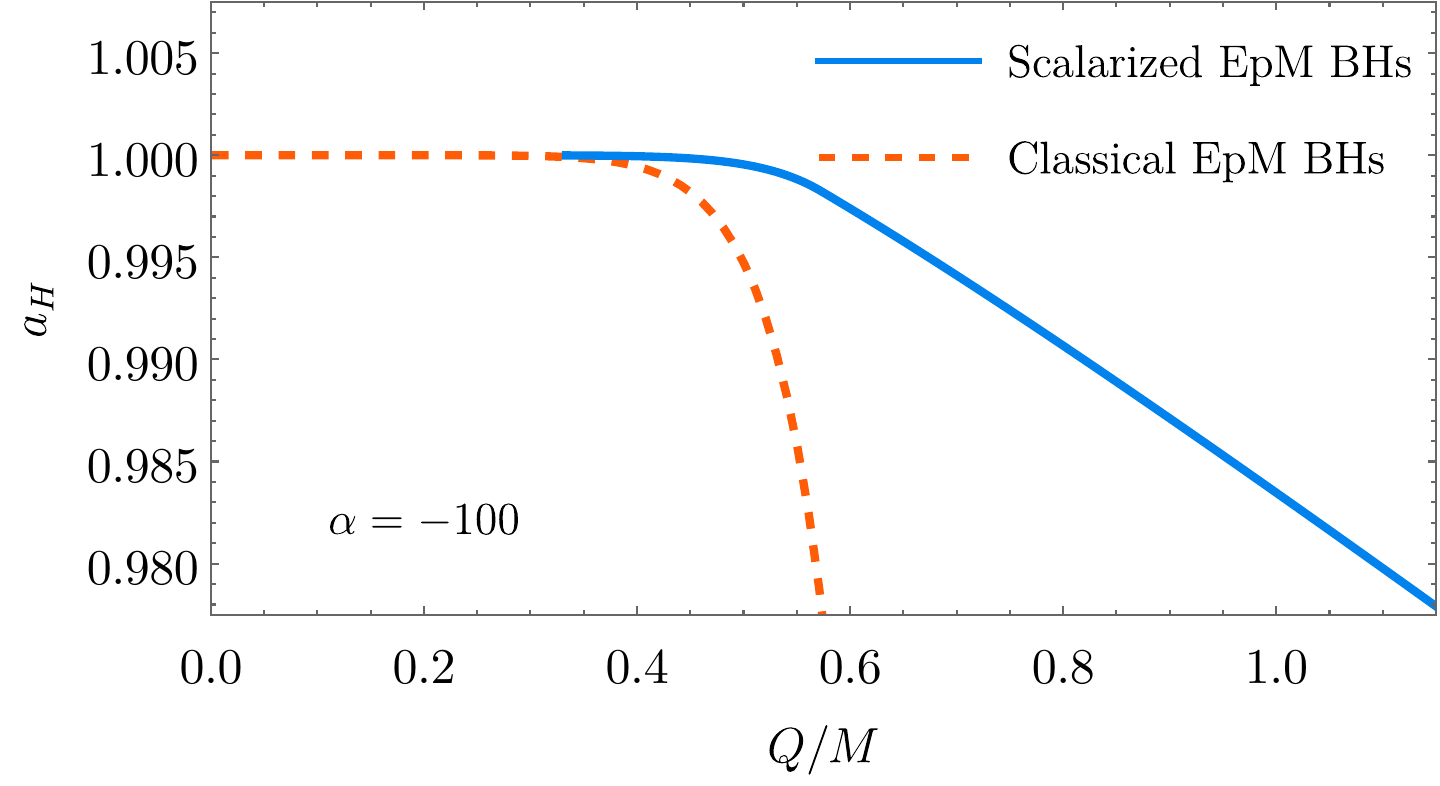}
    \caption{$a_H=A_H/(16\pi M^2)$ vs. $q=Q/M$ diagram, where $A_H$ is the horizon area of scalarized (solid line) and scalar--free (dashed line) EpM BHs with $\alpha=-100$.}
    \label{fig:aH}
\end{figure}

Note that there exists a region of non-uniqueness within the domain of existence where scalar--free and scalarized BHs coexist, as defined by the values of $q$ for which the metric $N$ possesses real roots. In this region, the scalarized solution is entropically preferred, as they maximize the entropy (or, equivalently, the horizon area $A_H$), as shown in \autoref{fig:aH} for the specific values indicated in the legend. It is worth highlighting that, for varying values of $\alpha$, the $a_H$-curves (as defined in the caption of~\autoref{fig:aH}) exhibit no significant deviation from one another.

\begin{figure}[h!]
    \centering
    \includegraphics[width=8.5cm]{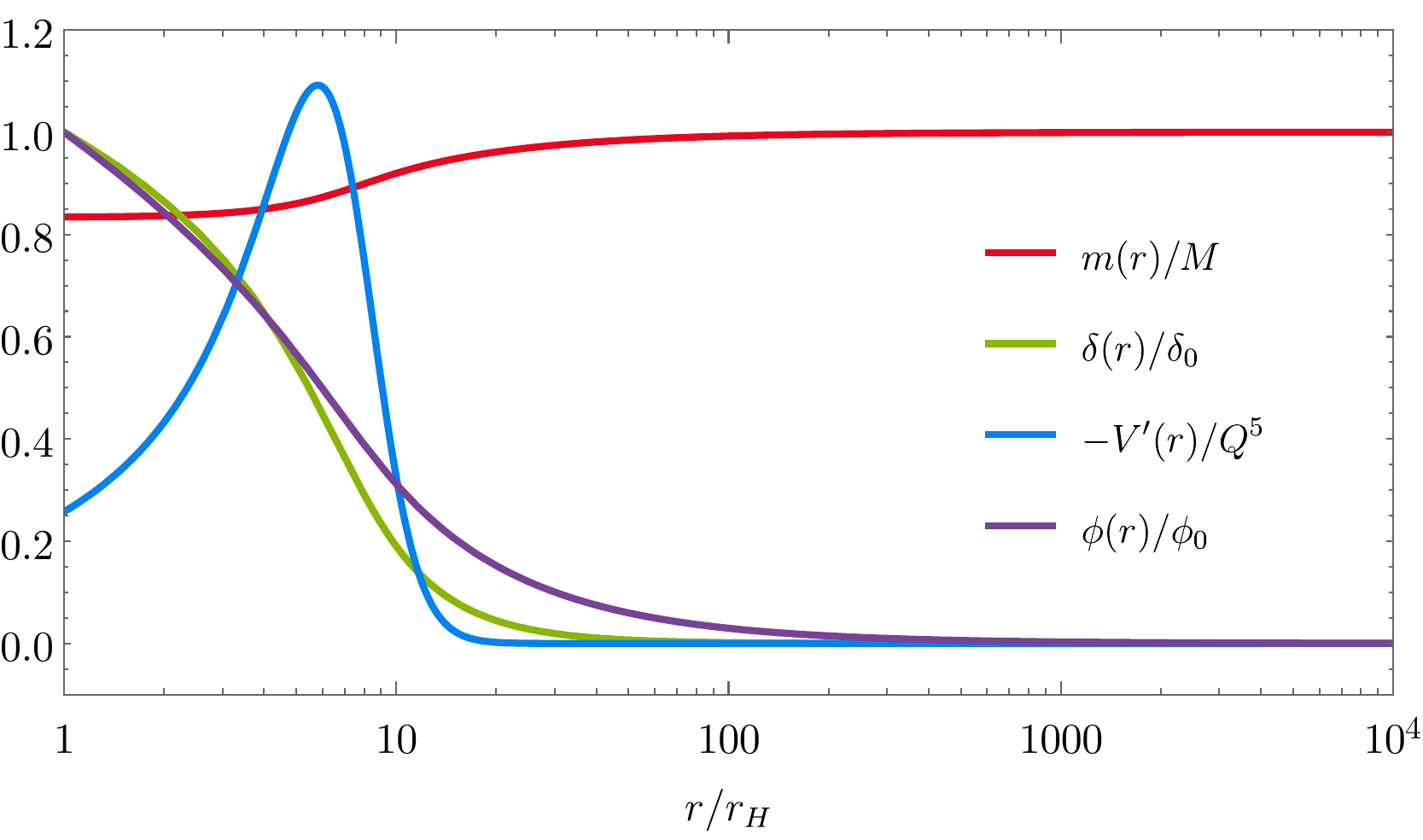}
    \caption{Metric and matter functions of a scalarized EpM BH with $Q=0.3$, $\alpha=-50$ and $r_H=0.1$.}
    \label{fig:solution}
\end{figure}

\begin{figure}[h!]
    \centering
    \includegraphics[width=\columnwidth]{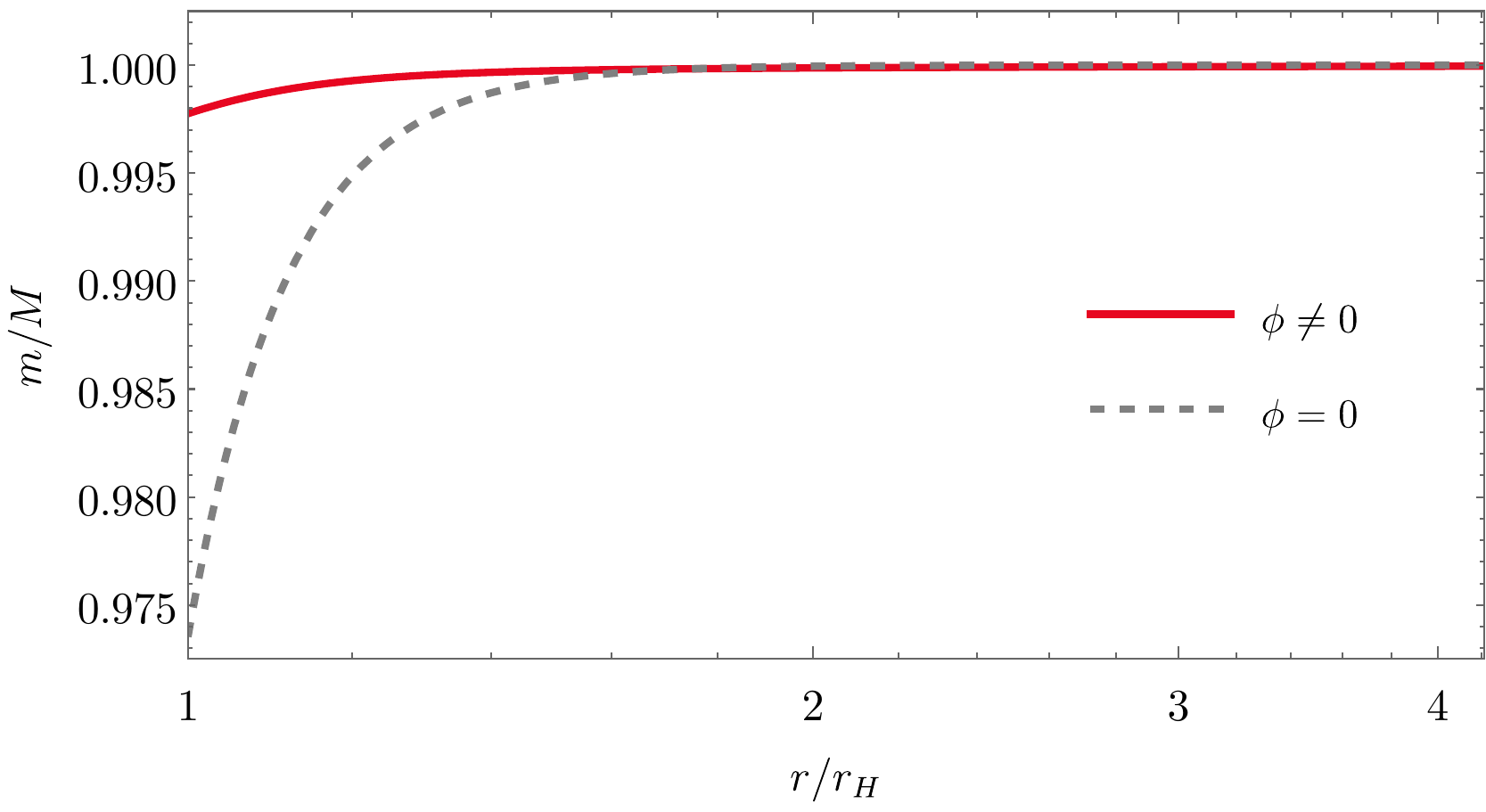}
    \caption{Mass function of a scalarized EpM and a scalar-free EpM BHs, with $Q=0.3$, $\alpha=-50$, $r_H^{\phi\neq 0}=0.95$ and $r_H^{\phi=0}=0.9177$.}
    \label{fig:mass}
\end{figure}

An illustrative solution is shown in \autoref{fig:solution} and \autoref{fig:mass} for  $Q=0.3$, $\alpha=-50$ and specific values of $r_H$ indicated in the legend. Note that the behavior of the whole functions coincides with the scalarized RN BH in \cite{Herdeiro:2018wub}. 

\section{Conclusions}
In this work, we analyzed the spontaneous scalarization undergone by an Einstein-power-Maxwell system non--minimally coupled to a scalar field. We obtained that, based on the virial identity, the power of the Maxwell scalar $F_{\mu\nu}F^{\mu\nu}$ cannot take any arbitrary value, but it must be constrained in a very specific way to ensure scalarization. We focused on the case $n<1$ and set the power in a way that the free solution corresponds to a straightforward modification of the RN metric, namely, a metric containing a term proportional to $r^{-\beta}$, with $\beta\in\mathbb{N}$. 
The motivation behind this choice was that considering $\beta$ as an integer is the most ``natural'' way to modify the mass function of the Schwarzschild BH, namely $m(r)=M-\xi/r^{\beta-1}$ ($\xi$ being a constant). For example, if $\beta=2$, the RN solution is recovered. If $\beta$ is arbitrary, the metric falls into the family of Kiselev solutions \cite{Kiselev:2002dx}. Framed in a power-Maxwell model, this modification of the mass function leads to certain values of $n$ that would not be as natural to choose if we did not have the guidance of the mass function (like $n=3/5$, in our case). Furthermore, our model shows advantages in comparison with the RN case because the domain of existence is significantly distinct from the $n=1$ scenario. Notably, the entire ($\alpha,q$)-range is expanded and the domain of the classical EpM BHs is extended, underscoring a clear advantage over the proposal outlined in ~\cite{Herdeiro:2018wub}.
Moreover, our findings demonstrated that the scalarized solutions are entropically favored in comparison to their hairless counterpart, in agreement with what occurs for the scalarized RN BHs in~\cite{Herdeiro:2018wub}. 
It could be interesting to explore arbitrary $(n,l,m)$ scalar clouds and to study the stability of the solution against perturbations. However, these and other aspects lie out of the scope of the present work and we leave them for future developments.

\section{Acknowledgements}
M.C expresses sincere gratitude to C. A. R. Herdeiro and the Gr@v Group at the University of Aveiro for their warm hospitality during the development of this work. This work is supported by the Center for Research and Development in Mathematics and Applications (CIDMA) through the Portuguese Foundation for Science and Technology (FCT -- Fundação para a Ciência e a Tecnologia), references \href{https://doi.org/10.54499/UIDB/04106/2020}{UIDB/04106/2020} and \href{https://doi.org/10.54499/UIDP/04106/2020}{UIDP/04106/2020}. The authors acknowledge support from the projects PTDC/FIS-AST/3041/2020 and CERN/FIS-PAR/0024/2021. This work has further been supported by the European Horizon Europe staff exchange (SE) programme HORIZON-MSCA-2021-SE-01 Grant No. NewFunFiCO-101086251. M.C acknowledges to the Master in Physics program at USFQ for financial support. N. M. S. is supported by the FCT grant SFRH/BD/143407/2019. 

\bibliographystyle{unsrt}
\bibliography{references.bib}


\end{document}